\documentclass[namedreferences]{solarphysics}

\usepackage[hyperref,optionalrh]{spr-sola-addons} 
\usepackage{graphicx}        
\usepackage{color}           
\usepackage{breakurl}        
\usepackage{multirow}
\usepackage{sidecap}

\chardef\us=`\_

\begin{document}

\begin{article}
\begin{opening}

\title{Forecasting Solar Cycle 25 using Deep Neural Networks\\ {\it Solar Physics}}

\author[addressref={aff1,aff2},email={bb0008@uah.edu}]{\inits{B.}\fnm{B.}~\lnm{Benson}\orcid{0000-0003-4945-3855}}
\author[addressref={aff1,aff2},corref,email={pand@uah.edu}]{\inits{W. D.}\fnm{W. D.}~\lnm{Pan}\orcid{0000-0001-7265-2188}}
\author[addressref={aff1,aff3},email={ap0162@uah.edu}]{\inits{A.}\fnm{A.}~\lnm{Prasad}\orcid{0000-0003-0819-464X}}
\author[addressref={aff1,aff3},email={allen.gary@uah.edu}]{\inits{G. A.}\fnm{G. A.}~\lnm{Gary}\orcid{0000-0002-7298-1405}}
\author[addressref={aff1,aff3},email={qh0001@uah.edu}]{\inits{Q.}\fnm{Q.}~\lnm{Hu}\orcid{0000-0002-7570-2301}}

\address[id=aff1]{The University of Alabama in Huntsville}
\address[id=aff2]{Department of Electrical and Computer Engineering}
\address[id=aff3]{Center for Space Plasma and Aeronomic Research}

\runningauthor{B. Benson \textit{et al.}}
\runningtitle{Forecasting SC-25 using Deep Neural Networks}

\begin{abstract}
With  recent advances in the field of machine learning, the use of deep neural networks for time series forecasting has become more prevalent. The quasi-periodic nature of the solar cycle makes it a good candidate for applying time series forecasting methods. We employ a combination of WaveNet and LSTM neural networks to forecast the sunspot number using the years 1749 to 2019 and total sunspot area using the years 1874 to 2019 time series data for the upcoming Solar Cycle 25. Three other models involving the use of LSTMs and 1D ConvNets are also compared with our best model. Our analysis shows that the WaveNet and LSTM model is able to better capture the overall trend and learn the inherent long and short term dependencies in time series data. Using this method we forecast 11 years of monthly averaged data for Solar Cycle 25. Our forecasts show that the upcoming Solar Cycle 25 will have a maximum sunspot number around 106 $\pm$ 19.75 and maximum total sunspot area around 1771 $\pm$ 381.17. This indicates that the cycle would be slightly weaker than Solar Cycle 24.  
\end{abstract}

\end{opening}

\section{Introduction}
     \label{S-Introduction} 

The solar cycle is a product of the solar dynamo processes that drive the Sun and is influenced by the cyclic regeneration of its magnetic field \citep{ref:t3, ref:t1}. It is quasi-periodic in nature and has a periodicity of approximately 11 years. The rise and fall in solar activity has a direct impact on the geospace environment and on life on Earth. An increase in solar activity comprises of increase in harmful EUV and X-ray emissions toward Earth which affect the temperature and density of our atmosphere. This can be harmful to the orbital lifetime of satellites in low Earth orbit \citep{ref:t2, ref:t1}. Higher solar activity which results in an increase in solar flares and coronal mass ejections (CMEs) can be harmful to communication systems, power systems, satellites and various other assets in addition to being harmful to astronauts in space. Therefore, it is in our best interest to have the ability to predict the strength of the solar cycle with good accuracy for long-term planning of space weather impacts on space missions and societal technologies.

Predicting the strength of the solar cycle is a non-trivial task due to the complexity of the solar dynamo. Forecasting strategies and methods have varied based on the type of indices used to estimate the strength and occurrence of the solar cycle. Statistics like periodicity and trends observed in previous cycles have been used in prediction of the solar cycle \citep{ref:t4, ref:t5, ref:t6}. Other indices were based on geomagnetic precursors \citep{ref:t7, ref:t8, ref:t9}, polar fields \citep{ref:t10, ref:t11, ref:t12} and flux transport dynamos \citep{ref:t13, ref:t14, ref:t15}. Neural networks trained on sunspot numbers have also been used in solar cycle predictions \citep{Fessant, ref:t29}. More recently, \citet{ref:t16} and \citet{ref:t17} used neural networks to predict the strength of the upcoming Solar Cycle 25.

The variation in the number of sunspots is an indicator of solar activity \citep{ref:t1}. It also directly corresponds with the total sunspot area that relates to the magnetic field entering the corona. The quasi-periodic nature of the solar cycle makes it a good candidate for applying time series forecasting methods to these datasets. Time series forecasting methods have been applied successfully to areas such as finance \citep{ref:t18}, meteorology \citep{ref:t19} and signal processing \citep{ref:t20}.

Classical methods for time series forecasting include models using moving averages (MA) and auto regression (AR) such as Autoregressive Moving Average (ARMA) and Autoregressive Integrated Moving Average (ARIMA), where the time series of historical observations is assumed to be linear and follow a known stochastic distribution. Other variants of these classical methods such as Seasonal Autoregressive Integrated Moving-Average (SARIMA), Seasonal Autoregressive Integrated Moving-Average with Exogenous Regressors (SARIMAX), Vector Autoregression (VAR) and Vector Autoregression Moving-Average (VARMA) have also been successfully applied \citep{ref:t26, ref:t27, ref:t28, sarima}. Recently, machine learning and deep-learning techniques have been extensively used on time series forecasting problems with better results. The deep-learning methods benefit from not having to assume any information about the long or short term distributions and from having the capability of model complex non-linear systems with rapidity. Studies have shown that neural network and deep-learning based models have outperformed these classical methods for time series forecasting tasks because of their ability to handle non-linearity which is more likely to be seen in real world problems \citep{ref:t21, ref:t22, ref:t16}.  

Recent research of convolutional networks with features such as dilated convolutions and residual connections outperform generic recurrent architectures for sequence modeling tasks \citep{conv_recurrent}. In this study we propose a model based on WaveNet \citep{ref:t23} which uses one-dimensional dilated convolutions, residual connections and LSTM \citep{ref:t24} that models the distribution of time series data with the capability to learn very long-term and short-term dependencies. We also do a comparison study of three other models using combination of LSTM and 1D ConvNets to find the best deep neural network model that is capable of delivering accurate forecasts. These models are applied on the monthly datasets of both observed sunspot number and total sunspot area with the goal of predicting the strength of the upcoming solar cycle. The objective of this paper is to show an improved method to forecast the next 11-year Solar Cycle 25. We present the data preparation, the applied neural network architecture, and the resulting predictions. 

\begin{figure}    
   \centerline{\includegraphics[width=0.45\textwidth,clip=]{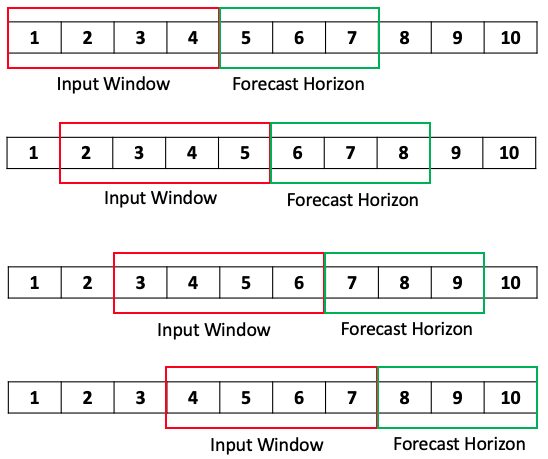}
              }
              \caption{Sliding window method illustrated with an example sequence of numbers from 1 through 10. Here each number represents one time step. The input window is slid one time step at a time throughout the whole sequence of data available to form four unique input and forecast horizon pairs (T=4, N=3).
                      }
   \label{time_window}
   \end{figure}
\section{Dataset Preparation}
    \label{Sec-2}
Predicting each of the sunspot number and sunspot area for the upcoming Solar Cycle 25 is cast as a univariate multi-step time series forecasting task. A historical time series $[x_1, y_2,\ .\ .\ . \ , x_T]$ is used as an input to predict the next N time steps $[x_{T+1}, x_{T+2}, . . . , x_{T+N}]$ known as the forecast horizon. The data is segmented using the sliding window method where a fixed window size of observations from the time series is chosen as an input and a fixed number of the following observations form the forecast horizon. This windowing process is repeated for the entire dataset by sliding the window one time step at a time to get the next slice of input and forecast horizon pairs.

\begin{figure}    
   \centerline{\includegraphics[width=1.25\textwidth,clip=]{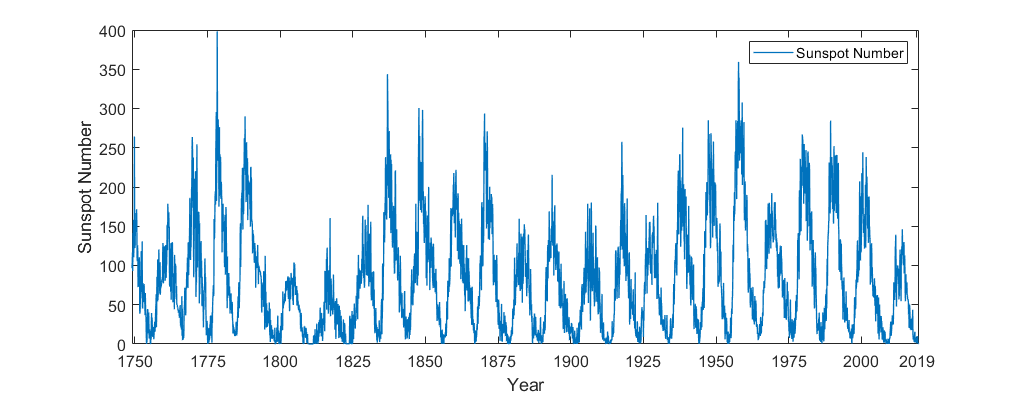}
              }
              \caption{Monthly averaged sunspot number for the years 1749 to 2019.
                      }
   \label{sunspot_number}
   \end{figure}

\begin{figure}    
   \centerline{\includegraphics[width=1.25\textwidth,clip=]{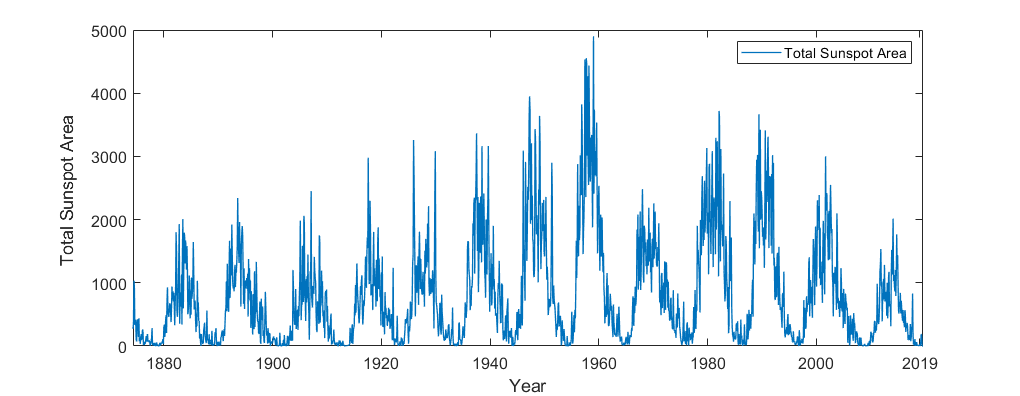}
              }
              \caption{Monthly averaged total sunspot area for the years 1874 to 2019.
                      }
   \label{sunspot_area}
   \end{figure}

\begin{figure}    
   \centerline{\includegraphics[width=0.85\textwidth,clip=]{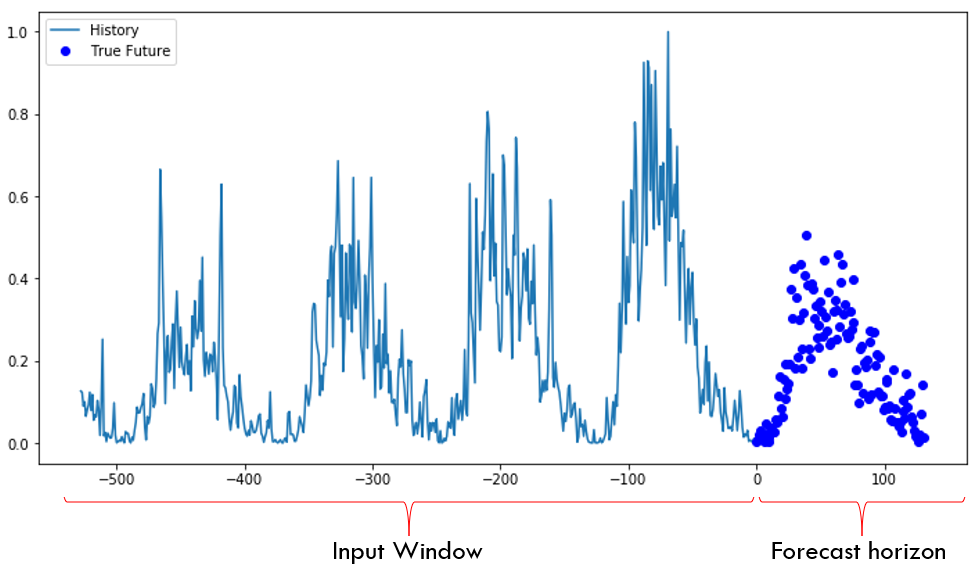}
              }
              \caption{Input window of size 528 time steps and forecast horizon of size 132 time steps for the sunspot area dataset.}
   \label{training_horizon}
   \end{figure}

\begin{figure}    
   \centerline{\includegraphics[width=1.25\textwidth,clip=]{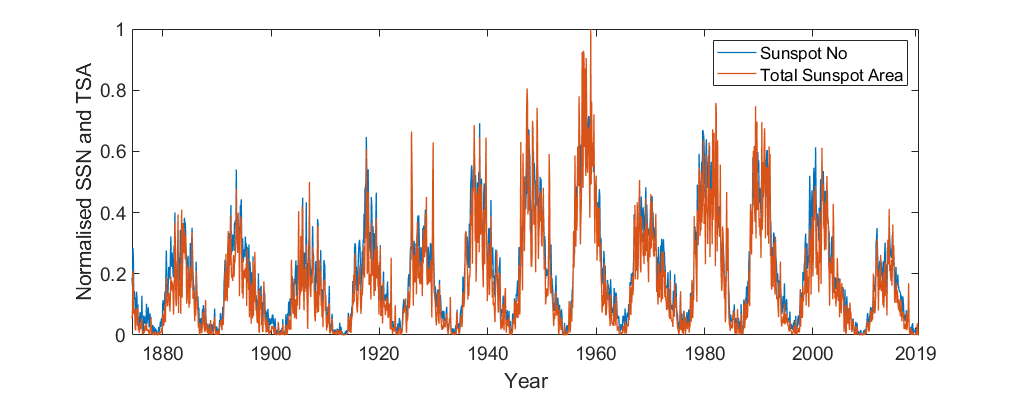}
              }
              \caption{Sunspot number (SSN) and total sunspot area (TSA) numbers normalized and aligned with the same dates between the years 1874 and 2019.
                      }
   \label{tsa_ssn}
   \end{figure}

Figure \ref{time_window} illustrates the sliding window method for the multi-step time series forecasting. The sunspot numbers were obtained from the World Data Center SILSO, Royal Observatory of Belgium, Brussels \url{http://sidc.be/silso/datafiles} \citep{sidc}. It contains 3251 records of monthly averaged sunspot number observations from the year 1749 to 2019. The sunspot area dataset is obtained from the website \url{http://solarcyclescience.com/activeregions.html} made available and maintained by Lisa Upton and David Hathaway. It contains daily sunspot area from the year 1874 to 2019. The daily data is averaged monthly and produces 1744 records. Both datasets are treated as a univariate time-series and the sliding window method is applied to them. In every alternate solar cycle the magnetic polarities of the sunspots in the northern and southern hemispheres change sign. This leads to at least two solar cycles of data being the minimum requirement needed to forecast the next cycle. However, two cycles are insufficient to notice the longer trend in data as seen in Figures \ref{sunspot_number} and \ref{sunspot_area}. By choosing the window size of four cycles we can ensure that the changes in polarity as well as the longer trend is noticeable and help in producing a more accurate forecast. Therefore, a  window size of 528 observations (4 cycles $\times$ 11 years/cycle $\times$ 12 months/year) and a forecast horizon of 132 observations (1 cycle $\times$ 11 years/cycle $\times$ 12 months/year) is chosen as seen in Figure \ref{training_horizon}. This produces 2560 unique input window and forecast horizon pairs for the sunspot number dataset.  The same sliding window method with a window size of 528 and forecast horizon size of 132 observations is applied to the monthly averaged sunspot area dataset and produces 1085 unique input window and forecast horizon pairs. As deep neural network models require large datasets, these input and forecast pairs are useful for providing accurate forecasts. Figure \ref{tsa_ssn} shows that when the sunspot number and total sunspot area datasets are normalized and aligned by the timeline, they show similar time variation. Thus, both datasets are indicative of each other in strength and time of occurrence. 

\begin{figure}    
   \centerline{\includegraphics[width=0.7\textwidth,clip=]{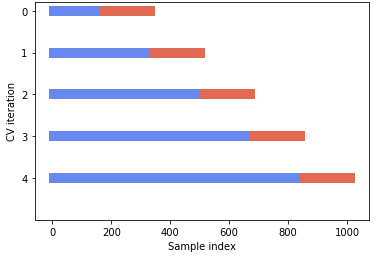}
              }
              \caption{``TimeSeriesSplit" cross-validation scheme using Scikit-learn. The sample index represents time steps (monthly data) for both SSN and TSA datasets. The blue represents the training set and orange represents the validation set.}
   \label{cross_validation}
   \end{figure}
   
In order to make an accurate forecast based on historical observations it is necessary for any model to be trained on the complete dataset. However, this makes it impossible to judge the performance of the model due to the absence of a ground truth to verify the forecast. Therefore, we divide our datasets into training and validation splits and pick the model with the best validation performance to be trained on the entire data to make our forecast. A time series cross-validation scheme known as ``TimeSeriesSplit"  is implemented from the widely used machine learning library Scikit-learn \citep{ref:t25}. The idea is to divide the training and validation sets at each fold or iteration such that the validation data is always ahead of the training data. Figure \ref{cross_validation} shows the cross-validation scheme \citep{cv}. This ensures that the chronological order is maintained thereby allowing the model to identify trends in data.

\section{Deep Neural Networks} 
      \label{S-general}      
Machine Learning methods such as Time Delay Neural Networks \citep{tdnn}, Support Vector Machines (SVM) \citep{ref:t30}, Random Forests \citep{ref:t31} and deep-learning methods such as Recurrent Neural Networks (RNN) , and Long Short-Term Memory (LSTM) \citep{ref:t24} have been extensively used for time series forecasting problems in domains such as finance, meteorology, statistics and signal processing. RNNs include an additional hidden state that passes on pertinent information learned from the current time step to the next time step, thereby allowing the model to learn the temporal dependencies in data. LSTMs are a type of RNN that include features such as memory gates and forget gates that make it possible to learn long-term dependencies. Studies have indicated that deep-learning models outperform stochastic models for time series forecasting problems \citep{ ref:t22, ref:t16}. Another method for processing sequential data such as time series data is by using one-dimensional (1D) convolutions. In a 1D convolution, each time step is obtained from a small patch or sub-sequence of temporal data in the input sequence. This extracted patch is then passed on through the neural network by adding weights and biases and produces a single time-step output. As the same input transformation is applied to each patch, patterns learned are easy to recognize at any position in the time sequence. 1D convnets can compete with RNNs delivering similar or better performance with much less training time.  More recently, WaveNet \citep{ref:t23} which is based on 1D dilated convolutions achieved state of the art performance in audio generation. Since audio data is also sequential, the same techniques can be applied to time-series forecasting.  For faster training and identifying longer trends in data, in this study, we propose a deep neural network model based on a combination of WaveNet and LSTM which include recurrent and 1D dilated methods.

\begin{figure}[h]    
   \centerline{\includegraphics[width=0.95\textwidth,clip=]{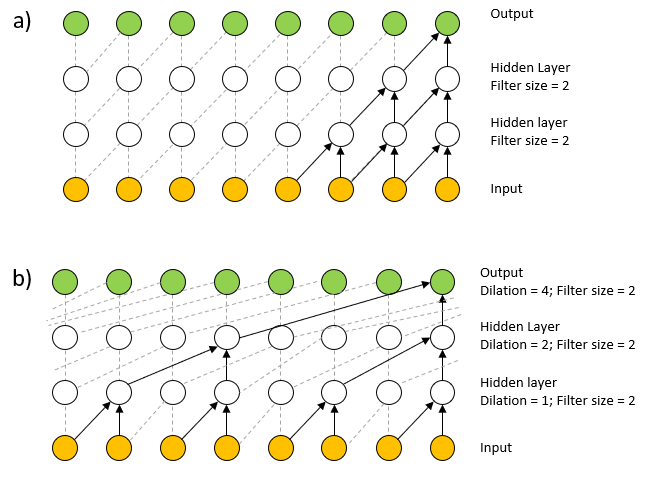}
              }
              \caption{a) A stack of 1D convolution layers. b) A stack of 1D dilated causal convolution layers using 1, 2 and 4 dilation rate.}
   \label{wavenet}
   \end{figure}
   
   \begin{figure}[h]    
   \centerline{\includegraphics[width=1\textwidth,clip=]{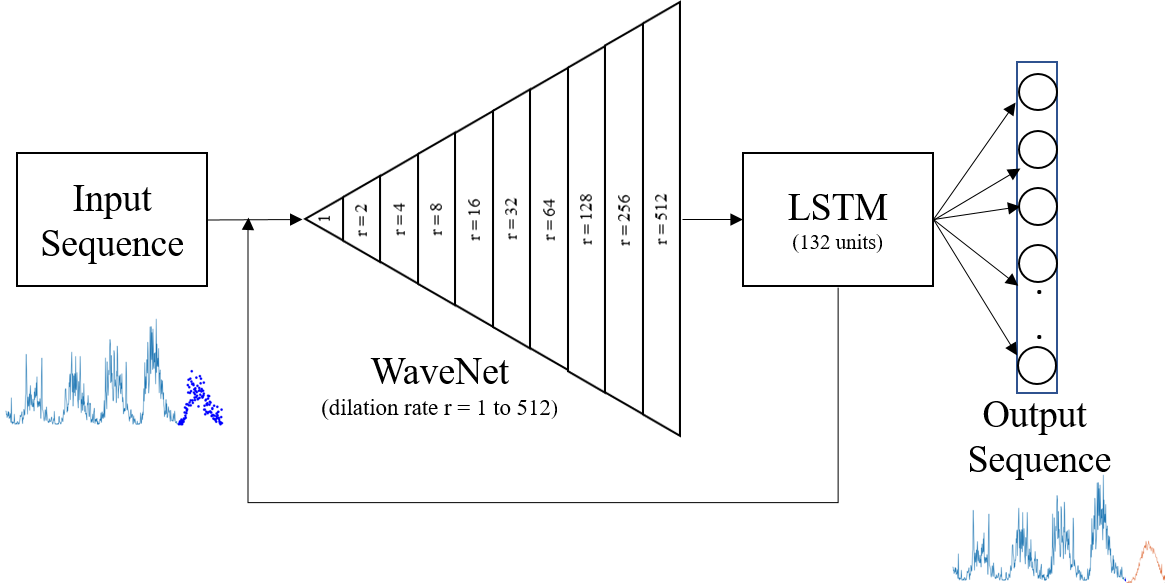}
              }
              \caption{ Proposed WaveNet + LSTM scheme showing the dilation rates, input and output sequences. }
   \label{wavenet_scheme}
   \end{figure}

\subsection{WaveNet Architecture} 
  \label{S-text}
Wavenet is an autoregressive generative model that operates on the time-series data directly. It learns to model the conditional probability distribution of the time-series data using a stack of 1D dilated causal convolution layers. For any given time series $x = \{x_{1},\ .\ .\ .\ ,\ x_{T}\}$ its joint probability is factorized as a product of its conditional probabilities.  

Equation \ref{Eq-1} shows the product of the conditional probabilities of the time series. Therefore, each time step $x_{t}$ is conditioned on the samples of all the preceding time steps. Dilated causal convolutions form the main ingredient of WaveNet models. They ensure that the model does not violate the order in which the data is modelled. In a dilated causal convolution layer filters are applied by skipping a constant dilation rate of inputs on the input sequence. The dilation rate is increased exponentially every layer which allows the model to have exponentially increasing receptive fields in each successive layer. 

   \begin{equation}  \label{Eq-1}
     p(x) = \prod_{t=1}^{T} p(x_{t}\ |\  x_{1},\ .\ .\ .\ ,\ x_{t-1})  \,.
   \end{equation}
   
Figure \ref{wavenet} shows both a stack of simple 1D convolution layers and a stack of dilated causal convolution layers as used in the WaveNet architecture. We can see for the dilated convolutions, with a dilation rate of 4 and filter size of 2 the output has a receptive field of 8 input units compared to only a receptive field view of only 4 units when dilations are not used. Therefore, by stacking a few layers of dilated convolutional layers we exponentially increase the receptive field allowing models using dilated convolutions to learn much longer sequences of time dependencies than traditional recurrent models \citep{conv_recurrent}. In addition to having a larger receptive field, WaveNets use gated activations, residual and skip connections in the neural network architecture. Gated activation units allow greater control than rectified activation whereas residual and skip connections enable faster convergence. In our proposed model shown in Figure \ref{wavenet_scheme}, we use a single LSTM layer at the end of the WaveNet model which ensures that our model learns both very long-term and shot-term time series dependencies from the input data.

\section{Experimental setup and Results} 
      \label{Experiments}      
In this study we compare the performance of four deep neural network models in order to find the model with the most accurate forecasts. The first model is a simple LSTM layer with 132 units. The second model consists of two stacked LSTM layers with 132 units each. The third model consists of a 1D convolution layer without any time dilations stacked with an LSTM layer of 132 units. The fourth model is the WaveNet architecture with dilation rates of 1, 2, 4, 8, 16, 32, 64, 128, 256 and 512 stacked with a single LSTM layer of 132 units. All four models are compared with a naive average forecast where all cycles are rescaled to have the same 132 month length and averaged over each time step. This provides us with a baseline to compare our models and measure the accuracy of the forecasts. To ensure unbiased results we use the same hyper-parameters such as dropout of 30$\%$ and batch normalization on each model. All experiments were performed on an NVIDIA DIGITS\textsuperscript{TM} workstation dedicated for deep-learning running on Ubuntu 16.04. It is equipped with four NVIDIA TITAN X GPUs with 12GB memory per GPU board and features 7 TFlops of single precision. All models were created using TensorFlow 2.0 and Keras using Python programming language \citep{ref:t32, ref:t33}. Root mean squared error (RMSE) was the chosen performance metric for all experiments to compare the performance of each model. The RMSE is the standard deviation of the residuals or prediction errors and provides a measure of how far the prediction is from the actual data. 

\begin{table}[h]
\caption{Cross-validation scheme using TimeSeriesSplit showing the different cross-validation (CV) folds with the number of training and validation pairs. 
}
\label{table-1}
\centering
\begin{tabular}{cccccc}
\hline
Dataset &
  Time steps &
  \begin{tabular}[c]{@{}c@{}}Input - Horizon\\  Pairs\end{tabular} &
  CV-Fold &
  \begin{tabular}[c]{@{}c@{}}Training \\ Sequence \\ Length\end{tabular} &
  \begin{tabular}[c]{@{}c@{}}CV \\ Sequence \\ Length\end{tabular} \\
 \hline
                                                              &      &      & 1 & 432  & 432 \\
                                                              &      &      & 2 & 864  & 432 \\
\begin{tabular}[c]{@{}c@{}}Sunspot\\ Number\end{tabular}      & 3251 & 2560 & 3 & 1296 & 432 \\
\textbf{}                                                     &      &      & 4 & 1728 & 432 \\
\textbf{}                                                     &      &      & 5 & 2160 & 432 \\
\textbf{}                                                     &      &      & 1 & 180  & 180 \\
\textbf{}                                                     &      &      & 2 & 360  & 180 \\
\begin{tabular}[c]{@{}c@{}}Total \\ Sunspot Area\end{tabular} & 1744 & 1085 & 3 & 540  & 180 \\
                                                              &      &      & 4 & 720  & 180 \\
                                                              &      &      & 5 & 905  & 180 \\
 \hline
\end{tabular}
\end{table}

\begin{figure}  
   \centerline{\includegraphics[width=1\textwidth,clip=]{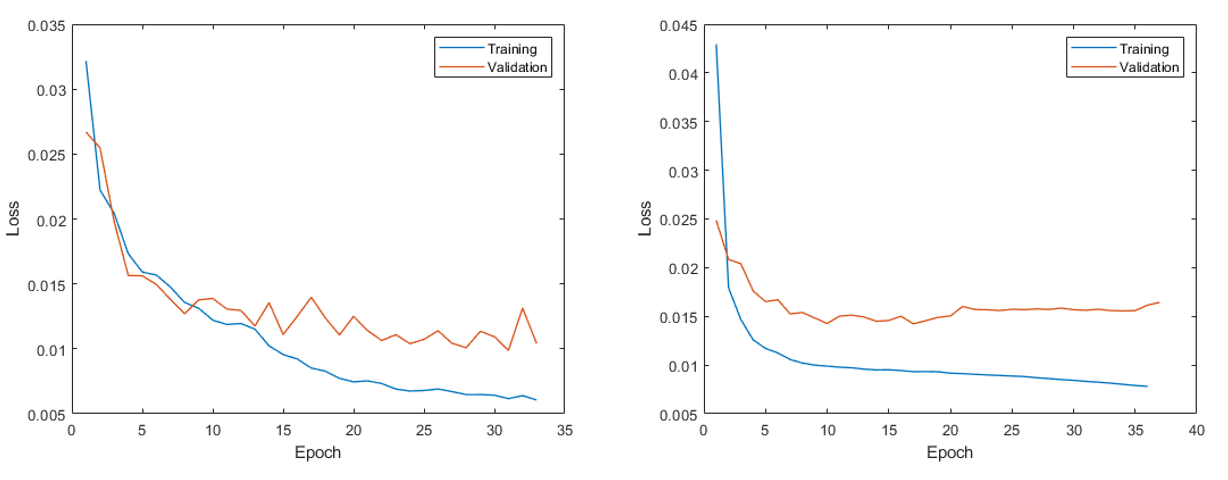}
              }
              \caption{Training and validation losses using the WaveNet + LSTM model for the sunspot number and sunspot area datasets respectively.}
   \label{loss}
   \end{figure}

The datasets for sunspot number and total sunspot area were processed as described in section \ref{Sec-2}. The data is also normalized to values between 0 and 1 to speedup the convergence of the gradients. To ensure there is enough data in the validation set we implement a 5-fold cross-validation scheme for both datasets using TimeSeriesSplit from the Scikit-learn library. Table \ref{table-1} describes the cross-validation scheme and the split used for training and validation data. The Adam optimization algorithm was used to update weights while training on sequences of batch size 32 with a learning rate of $5\times10^{-4}$ and a decay rate of $10^{-6}$ for 100 iterations. This allowed for faster convergence of the loss function. The tensorflow.keras library allows for a scheme called  ``early-stopping" from the ``callbacks" module to avoid overfitting where we can stop the training when the model starts to overfit with a ``patience" value (set to 5 iterations). This gives control over the number of times the validation loss is allowed to exceed its previous best value. We also used this scheme to save the model and its weights at a checkpoint with the best performance. Figure \ref{loss} shows the training and validation loss curves for the sunspot number and total sunspot area datasets. The curves show that the model stopped training before reaching 100 iterations for both datasets.  

\begin{table}[]
   
\caption{Performance summary of all models on the sunspot number and total sunspot area datasets.  
}
\label{table-2}
\begin{center}
\centering
\begin{tabular}{lll}
\hline
Dataset            & Model          & RMSE  \\
\hline
Sunspot Number    & Baseline (Average)           & 34.15   \\
				   & LSTM           & 4.42   \\
                   & Stacked LSTM   & 4.09    \\
                   & 1D Conv + LSTM & 3.89   \\
                   & \textbf{WaveNet + LSTM} & \textbf{2.93}  \\
Total Sunspot Area  & Baseline (Average)           & 481.56   \\
					& LSTM           & 13.41 \\
                   & Stacked LSTM   & 13.99 \\
                   & 1D Conv + LSTM & 12.70  \\
                   & \textbf{WaveNet + LSTM} & \textbf{11.11} \\
\hline
\end{tabular}
\end{center}
\end{table}

\begin{figure}    
   \centerline{\includegraphics[width=1\textwidth,clip=]{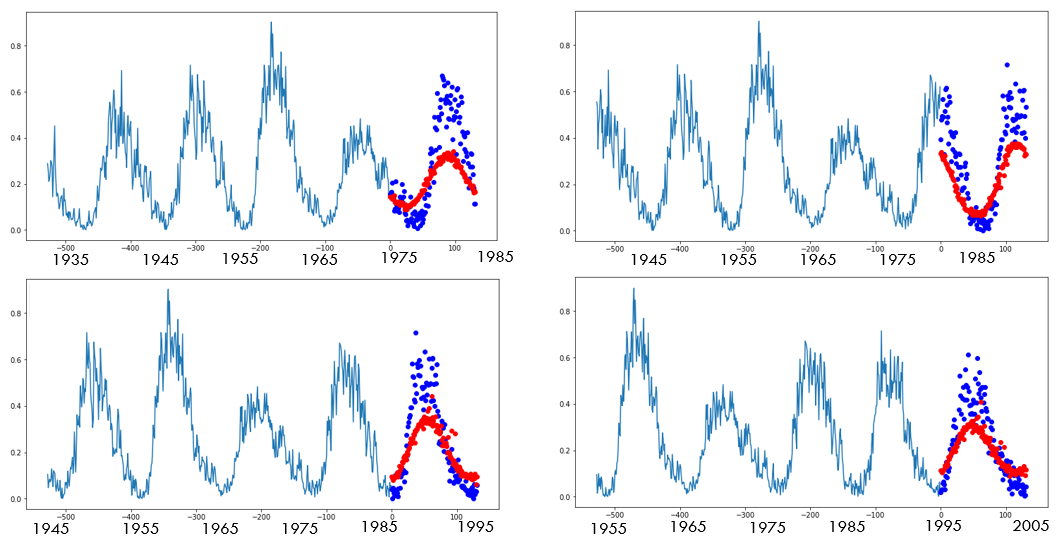}
              }
              \caption{Forecast graphs for the sunspot number dataset for the Stacked LSTM model. ``History" is represented by the solid blue, ``True future" by the blue dots and ``Forecast" by the red dots. }
   \label{stacked_lstm}
   \end{figure}

\begin{figure}    
   \centerline{\includegraphics[width=1\textwidth,clip=]{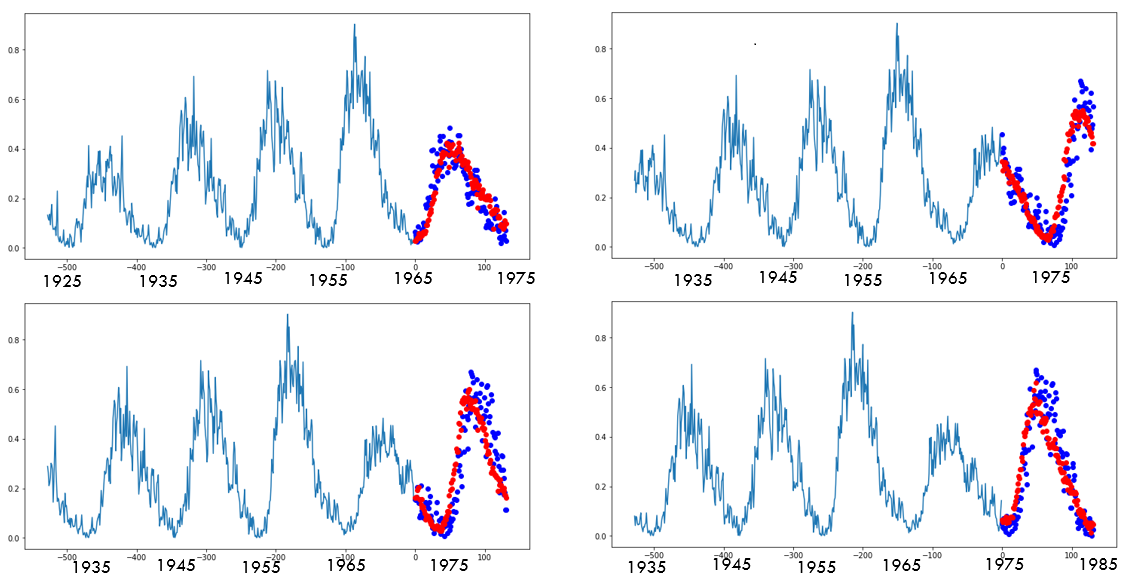}
              }
              \caption{Forecast graphs for the sunspot number dataset for the WaveNet + LSTM model. ``History" is represented by the solid blue,``True future" by the blue dots and ``Forecast" by the red dots. }
   \label{ssn_val}
   \end{figure}

\begin{figure}    
   \centerline{\includegraphics[width=1\textwidth,clip=]{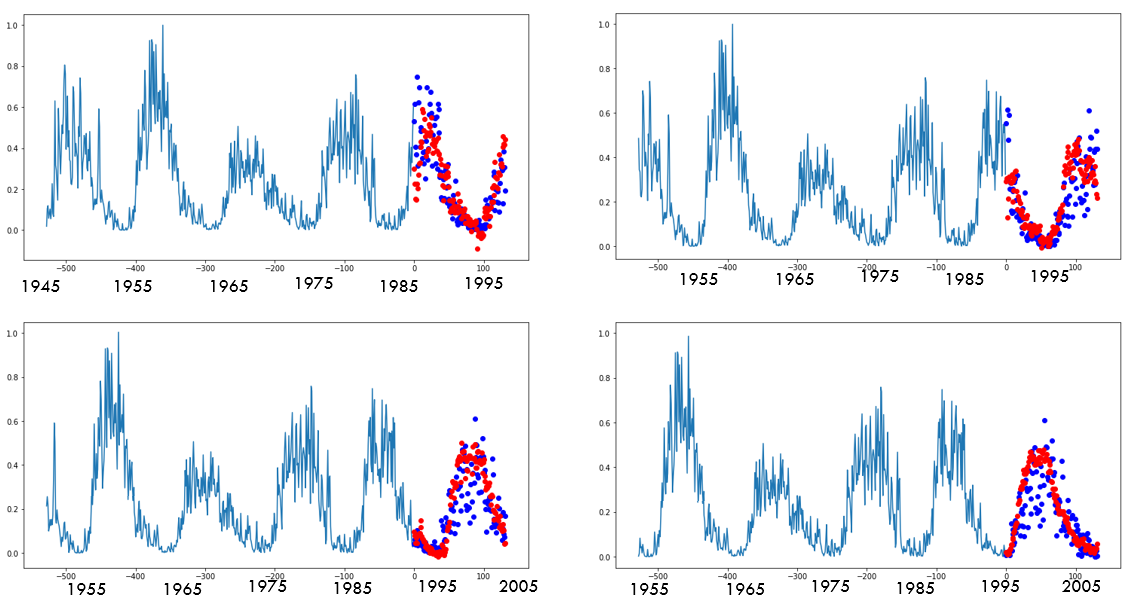}
              }
              \caption{Forecast graphs for the total sunspot area dataset for the WaveNet + LSTM model. ``History" is represented by the solid blue, ``True future" by the blue dots and ``Forecast" by the red dots. }
   \label{tsa_val}
   \end{figure}

Table \ref{table-2} summarizes the performance of all models on the sunspot number and total sunspot area. The WaveNet + LSTM model performed the best on both datasets. Figures \ref{stacked_lstm}, \ref{ssn_val} and \ref{tsa_val} show the true history and forecast of the sunspot and total sunspot area datasets for the Stacked LSTM and WaveNet + LSTM models. The figures show training data used as ``History", the actual forecast horizon as ``True Future" and the predicted forecast as ``Forecast". From Figures \ref{stacked_lstm} and \ref{ssn_val} it can be seen that the predicted forecasts for the WaveNet + LSTM model perform far better than the Stacked LSTM model on the sunspot number dataset even though there is comparatively small difference in their RMSE values. This reflects the ability of WaveNets to learn very long-term and well as short-term dependencies from the input data. The RMSE values for the total sunspot area dataset across all models is higher than the models for the sunspot number dataset. This is expected due to the limited amount of data available for the total sunspot area. To verify this, we extracted data from the sunspot area dataset to match the same time period of the total sunspot area dataset and trained it using our WaveNet + LSTM model which resulted in an RMSE value of 8.51 that is closer to the RMSE value for the total sunspot area dataset.

 \begin{figure}    
   \centerline{\includegraphics[width=1\textwidth,clip=]{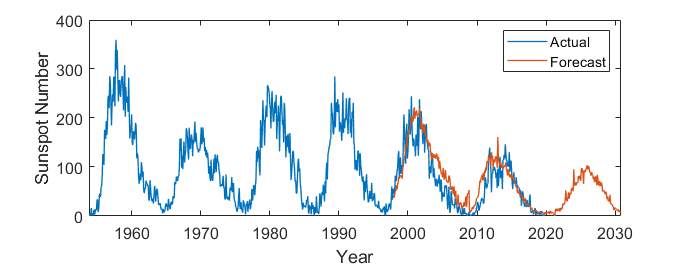}
              }
              \caption{Actual vs forecast graph for the sunspot validation data using WaveNet + LSTM model showing the predictions for the last two solar cycles and Solar Cycle 25.}
   \label{final_val_ssn}
   \end{figure}

    \begin{figure}    
   \centerline{\includegraphics[width=1\textwidth,clip=]{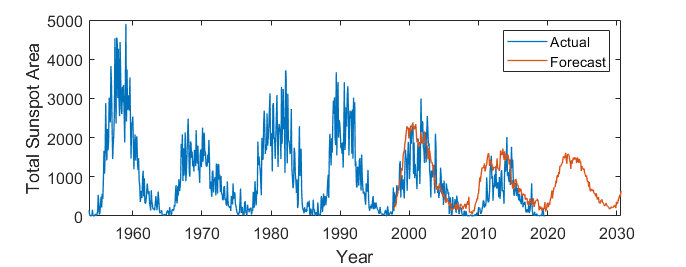}
              }
              \caption{Actual vs forecast graph for the total sunspot area validation data using WaveNet + LSTM model showing the predictions for the last two solar cycles and Solar Cycle 25.}
   \label{final_val_tsa}
   \end{figure}

\begin{figure}    
   \centerline{\includegraphics[width=1\textwidth,clip=]{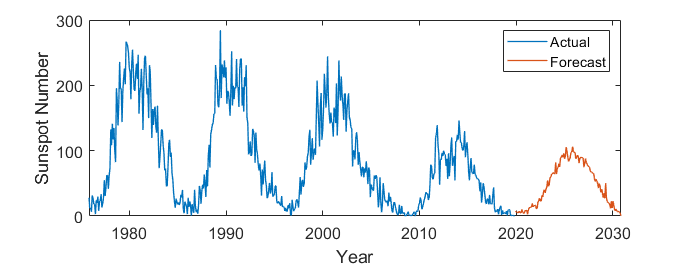}
              }
              \caption{Solar Cycle 25 forecast for the sunspot number data. }
   \label{final_ssn}
   \end{figure}

\begin{figure}   
   \centerline{\includegraphics[width=1\textwidth,clip=]{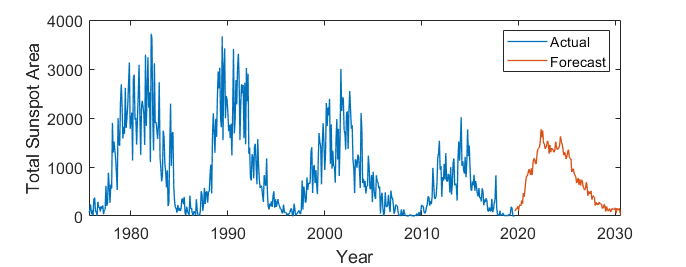}
              }
              \caption{Solar Cycle 25 forecast for the total sunspot area data.}
   \label{final_tsa}
   \end{figure}

As discussed in section \ref{Sec-2}, we chose the model with the best validation performance to train on the whole dataset to produce our 11-year forecast for the upcoming Solar Cycle 25. Figures \ref{final_val_ssn} and \ref{final_val_tsa} show the actual vs. forecast predictions for the sunspot number and total sunspot area using validation data with the WaveNet + LSTM model. The forecasts for Solar Cycle 23 and Solar Cycle 24  show that the model is able to predict the trends in data and also forecast the strength of those cycles accurately for both datasets, although, the total sunspot area shows a time lag of about 1.5 years.  Figures \ref{final_ssn} and \ref{final_tsa} show the forecast for the upcoming Solar Cycle 25 using the entire data for training the WaveNet + LSTM model. Both cycles show similar strength and suggest that Solar Cycle 25 will be slightly weaker compared to the previous Solar Cycle 24. The forecast for the sunspot numbers dataset suggest a peak of 106 $\pm 19.75$ with the peak occurring in March 2025, whereas, the forecast for the total sunspot area suggest a peak of 1771 $\pm 381.17$ with the cycle reaching its peak in May 2022. The discrepancy in the peak dates is expected as seen in the validation data forecasts. However, as stated earlier, both forecasts suggest similar strength of the solar cycle. Using the mean average error we determine that the uncertainty in predictions is 8\% for the sunspot number dataset and 12\% for the total sunspot area dataset. 

Forecasts for Solar Cycle 25 have varied from suggesting a stronger cycle than Solar Cycle 24 to the weakest cycle ever recorded with sunspot numbers ranging from 57-167. \citet{ref:t17} used spatial-temporal data with neural networks to predict that the upcoming Solar Cycle 25 would be the weakest cycle ever recorded with sunspot numbers of 57{$\pm$} 17 and total sunspot area of $\approx$ 700 with a peak around 2022-2023. \citet{hathaway2018} used a flux transport model and predicted that Solar Cycle 25  would be similar in size to Solar Cycle 24 with a 15{\%} uncertainty. \citet{Labonville2019} used a dynamo-based model to forecast the upcoming solar cycle and predicted a maximum sunspot number of 89 +29/-15. An international panel co-chaired by NOAA/NASA released a preliminary forecast on April 5, 2019 with the consensus predicting that Solar Cycle 25 would be similar in size to Solar Cycle 24. The minimum and maximum sunspot number forecast were 95 and 130 respectively.  \citet{ref:t16} used two layers of stacked LSTMs and predicted that the upcoming Solar Cycle 25 would have a maximum sunspot number of 167.3 with the peak being reached in 2023.2 {$\pm 1.1$}. The low RMSE values of the model and its forecasts on Solar Cycle 23 and Solar Cycle 24 give us confidence that our model is the best deep neural network based approach using temporal indices of sunspot numbers and total sunspot area for predicting the strength of the upcoming solar cycle. 
   
\section{Conclusion} 
      \label{S-Conclusion} 
In this study, we presented four deep neural network models with the goal of predicting the strength of the upcoming Solar Cycle 25 using monthly averaged data from the sunspot number and total sunspot area datasets. We clearly demonstrated that the proposed WaveNet + LSTM model performed best compared to the other deep neural network based models. It is capable of modeling both long-term and short-term dependencies and of identifying trends in time series data as seen in forecasts in the validation data along with forecasts of cycles  Solar Cycle 23 and  Solar Cycle 24. Our forecasts indicate that Solar Cycle 25 will be slightly weaker than Solar Cycle 24 with a maximum sunspot number of 106 $\pm$ 19.75  with 8\% uncertainty and total sunspot area of 1771 $\pm$ 381.17 with 12\% uncertainty and the cycle reaching its peak in May 2025 ($\pm$ one year). Our forecast falls within the uncertainty of \citet{hathaway2018} and the NOAA/NASA forecasts. This is consistent with the consensus forecast of a weak cycle ahead.   

Our proposed method can be applied to any univariate time series data that exhibits properties such as trend and seasonality. One limitation is we cannot expect to forecast time series data too far into the future. When we feed a period of forecast as input back into the model the errors tend to accumulate and get worse over time. This work can also be extended to including other related parameters that are indicative of the solar cycle and forecast them as a multivariate time series. We hope that this study would encourage the use of deep neural networks in forecasting tasks in heliophysics.     

\begin{acks}
 The authors would like to thank Dr. David Hathaway and Dr. Lisa Upton for maintaining and providing us with the data for the total sunspot area.
\end{acks}

\bigskip
\begin{footnotesize}
\noindent \textbf{Disclosure of Potential Conflicts of Interest} \quad The authors declare that they have no conflicts of interest.
\end{footnotesize}



\bibliographystyle{spr-mp-sola}
\bibliography{bibliography}  

\IfFileExists{\jobname.bbl}{} {\typeout{}
\typeout{****************************************************}
\typeout{****************************************************}
\typeout{** Please run "bibtex \jobname" to obtain} \typeout{**
the bibliography and then re-run LaTeX} \typeout{** twice to fix
the references !}
\typeout{****************************************************}
\typeout{****************************************************}
\typeout{}}

\end{article} 

\end{document}